\documentclass[aps,prl,twocolumn,groupedaddress,showpacs]{revtex4}
\usepackage[dvips]{graphicx,color}
\begin{document}

\title{Atom detection and photon production in a scalable, open, optical microcavity}

\author{M. Trupke}
\author{J. Goldwin}
\author{B. Darqui{\'e}}
\author{G. Dutier}
\altaffiliation{Present address: UMR 7538 du CNRS, Universit\'{e}
Paris 13, 99 Av J.-B. Cl\'{e}ment, 93430 Villetaneuse, France.}

\author{S. Eriksson}
\author{J. Ashmore}
\author{E. A. Hinds}

\affiliation{Centre for Cold Matter, Imperial College, Prince
Consort Road, London SW7 2BW, United Kingdom}
\date{\today}

\begin{abstract}
A microfabricated Fabry-Perot optical resonator has been used for
atom detection and photon production with less than 1 atom on
average in the cavity mode. Our cavity design combines the intrinsic
scalability of microfabrication processes with direct coupling of
the cavity field to single-mode optical waveguides or fibers. The
presence of the atom is seen through changes in both the intensity
and the noise characteristics of probe light reflected from the
cavity input mirror. An excitation laser passing transversely
through the cavity triggers photon emission into the cavity mode and
hence into the single-mode fiber. These are first steps towards
building an optical microcavity network on an atom chip for
applications in quantum information processing.
\end{abstract}

\pacs{42.50.Pq, 03.67.Lx, 03.75.Be}
\maketitle

When a neutral atom is placed in a high-finesse optical cavity, the
electric dipole coupling between the atom and the light field can
lead to quantum coherence between the two. This fact forms the basis
of cavity quantum electrodynamics (QED)~\cite{Berman1994}. Recently,
there has been considerable interest in the possibility of applying
cavity QED to problems in quantum information processing, as
reviewed, for example, in Ref.~\cite{vanEnk2004}. Single photons
have been generated on demand from falling \cite{Kuhn2002} and
trapped~\cite{McKeever2004} atoms in high-finesse Fabry-Perot
cavities, and recent experiments have investigated the
cavity-assisted generation of single photons from atomic
ensembles~\cite{Vuletic}. These are important steps towards building
multiple-cavity quantum information networks, such as those proposed
in Ref.~\cite{Cirac1997}. However, experiments so far have been
limited to single cavities by the technical demands of achieving
high enough finesse. Outstanding challenges now are to make the
cavities smaller, to fabricate them in large numbers with the
possibility of multiple interconnects, and to load them conveniently
and deterministically with atoms. This would pave the way to
circuit-model quantum computers~\cite{DiVincenzo2000}, to one-way
computations based on cluster states~\cite{clusters}, and to other
schemes requiring multiple cavities~\cite{multicavity}.

As a first move in this direction, two recent experiments have used
a small magnetic guide to load atoms into a cavity
~\cite{Teper2006}. However the cavities in these experiments were
$2$-$3\,\mbox{cm}$ long and therefore not more scalable than a
conventional Fabry-Perot cavity. By contrast, Aoki {\it et al.}\
have dropped atoms close to a microscopic toroidal cavity and
observed evidence of strong coupling~\cite{Aoki2006}. These
resonators can be microfabricated in large arrays, however they are
not easily used for controlled atom-cavity coupling because of the
need to position the atom very precisely in the evanescent field
just outside the surface of the resonator. For this reason it is of
interest to consider microscopic Fabry-Perot cavities, whose open
structure gives access to the central part of the cavity field. In
one recent design~\cite{Steinmetz2006}, the two mirrors of such a
resonator are formed by optical fibers whose tips have been modified
into concave reflectors. This design can achieve small mode volumes
and can collect the light efficiently, but is not easily scaled.

In collaboration with the group of M. Kraft, we have developed a
unique resonator design using curved micromirrors fabricated at
desired positions on a silicon wafer~\cite{Eriksson2005,Trupke2005}.
The cavities are closed by plane mirrors on the ends of optical
fibers or waveguides. This design combines the intrinsic scalability
of microfabrication processes with direct coupling of the cavity
field to single-mode optical waveguides or fibers. Advances in the
production of etched waveguides~\cite{microoptics} and plane-ended
fiber arrays, together with the recent production of etched
three-dimensional actuators~\cite{Gollasch2005}, promise to make
these cavities fully integrated and scalable. Here we report on the
first use of such cavities to detect atoms and produce photons. We
study the reflection spectrum and noise characteristics of the
coupled atom-cavity system and we trigger the cavity-enhanced
emission of a few photons from the atoms into the fiber. These are
all essential steps towards integrating the cavities on an atom chip
for quantum information processing.

The paraxial TEM00 cavity mode with wavenumber $k$ has an intensity
distribution given by $D(\rho,z)=(w_0/w)^2
\sin^2(kz)\exp(-2\rho^2/w^2)$, where $w$ characterizes the
transverse size of the mode, $w_0$ is its minimum value, and
$\rho,z$ are the cylindrical coordinates. The integral of this
distribution over a cavity of length $L$ gives the mode volume
$V=\pi w_0^2 L/4$, and the weighted sum over the positions of the
atoms $\sum_jD(\rho_j,z_j)\equiv N$ defines the effective number of
atoms in the cavity. For an atom placed at the maximum of $D$, the
coherent atom-cavity coupling is characterized by $g_0=
\mu\sqrt{\omega_C/(2\hbar\,\varepsilon_0V)}$, where $\mu$ is the
electric dipole transition moment, $\omega_C$ is the angular
frequency of the mode, and $\varepsilon_0$ is the free-space
permittivity. Our microfabricated cavities give large values of
$g_0$ by virtue of their extremely small mode volume. For example,
the experiments described here use a cavity with $L=130\,\mu$m and
$w_0=4.6\,\mu$m. This gives $g_0=6.1\times 10^8\,\mbox{s}^{-1}$ for
$\sigma^\pm$-polarized light driving the cycling transition
$\left|F=3,M=\pm 3\right\rangle \to \left|F^\prime=4,M^\prime=\pm
4\right\rangle$ of the D$_2$ line in $^{85}$Rb, which is used in
this work. This rate is equal to half the single-photon Rabi
frequency. It is to be compared with $\gamma=1.9\times
10^7\,\mbox{s}^{-1}$~\cite{Volz1996}, which is half the natural
decay rate of the excited atomic population and with $\kappa$, which
is half the decay rate of power in the cavity. The rate $\kappa$ is
related to the cavity finesse $\mathcal{F}$ by $\kappa=\pi
c/(2L\mathcal{F})$ ($c$ is the speed of light), and hence to the
reflectivities of the mirrors~\cite{SalehTeich}. Although we have
previously made microcavities with finesse $\mathcal{F}>5000$
\cite{Trupke2005}, these first experiments with atoms use
$\mathcal{F}=280$, corresponding to $\kappa=1.3\times
10^{10}\,\mbox{s}^{-1}$. From these three rates, one can construct
the dimensionless parameter $\kappa\gamma/g_0^2$, which is the
number of atoms needed to modify the cavity field
appreciably~\cite{Purcell,Berman1994}. This has the value $0.7$ for
this work, indicating that a single atom is enough to significantly
effect the cavity field.

\begin{figure}
\includegraphics[scale=0.6]{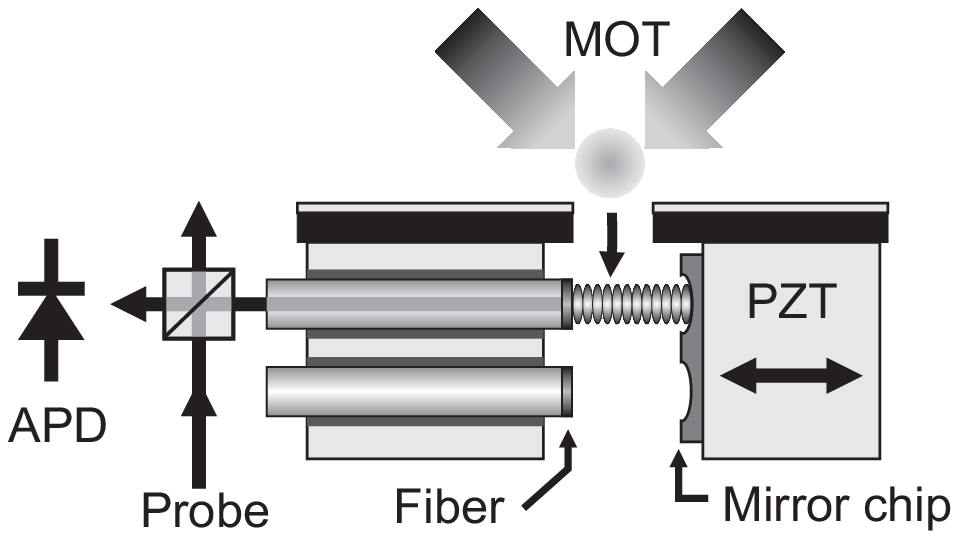}\\
\includegraphics[scale=0.55]{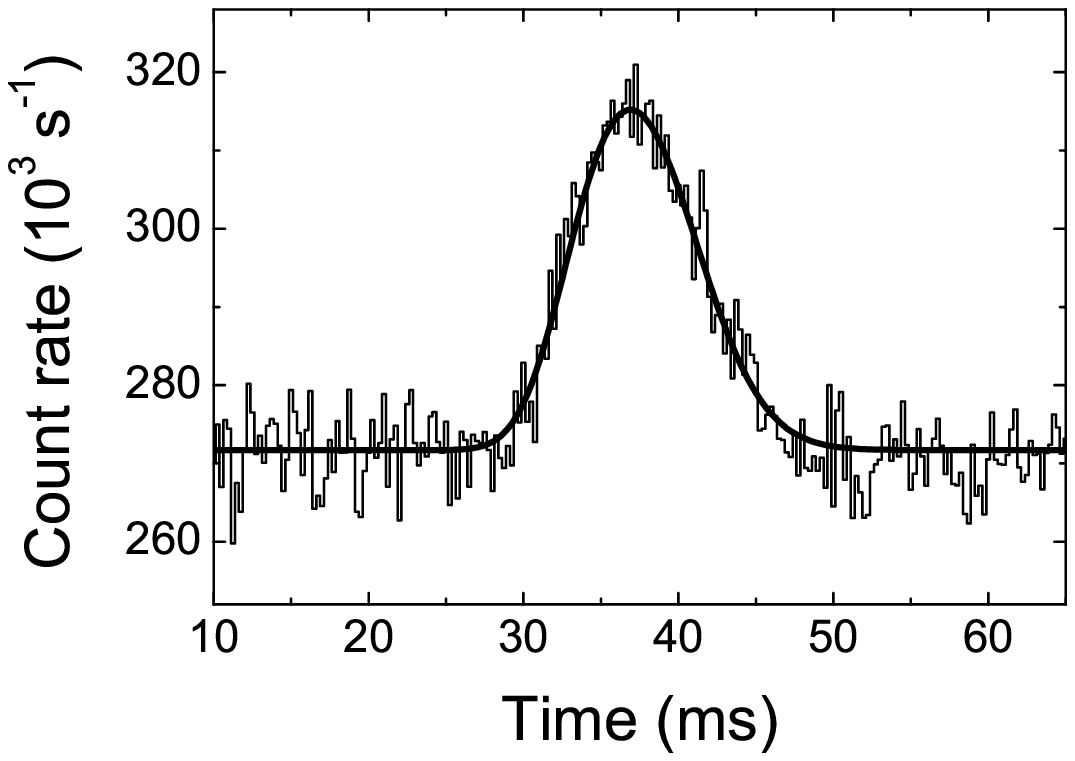}
\caption{\label{fig:expt}Apparatus (not to scale). Graph: change in
reflected probe light as atoms traverse the cavity. Averaging
window: $250\,\mu$s. Laser, cavity and atom frequencies are equal:
$\omega_L=\omega_C=\omega_A$. Atoms are released from optical
molasses at time $t=0$, and 34 identical drops are averaged. There
are $\sim0.7$ atoms in the cavity on average at the peak.}
\end{figure}

The apparatus is shown in Fig.~\ref{fig:expt}. Two fibers with plane
dielectric mirrors on their ends are held in parallel grooves
$2\,\mbox{mm}$ apart on a glass-ceramic substrate. The fibers face
two of an array of concave mirrors etched into a silicon wafer on a
$500\,\mu\mbox{m}$ square grid. These mirrors have a
$186\,\mu\mbox{m}$ radius of curvature and $100\,\mu\mbox{m}$
diameter, and they are coated with a dielectric multilayer to
reflect light at $780\,\mbox{nm}$. The silicon mirror chip is
mounted on a translator (PZT), which is adjusted every
$100\,\mbox{ms}$ under computer control to tune the upper cavity to
the free-space atomic resonance. Both cavities can be run
simultaneously, but in this work the lower cavity was not used. A
mirror is used to form a magneto-optical trap (MOT) $6\,\mbox{mm}$
above the active cavity, where up to $4\times 10^7$ cold $^{85}$Rb
atoms are collected. The magnetic quadrupole field of the MOT is
switched off and optical molasses~\cite{subdoppler} cools the cloud
over $15\,\mbox{ms}$ to $\sim 30\,\mu$K. The light is then shut off,
allowing the atoms to fall into the microcavity through a
$1\,\mbox{mm}$ hole in the mirror. Weak probe light (typically $\sim
1$ pW) is incident on the upper cavity through a beamsplitter and
the reflected photons are counted in $10\,\mu$s bins by a Perkin
Elmer SPCM-AQR-14 avalanche photodiode (APD) to detect the presence
or absence of atoms. For the experiment presented in
Fig.~\ref{fig:expt}, probe light reflected from the cavity input
mirror produces an APD count rate far from resonance of
$I_0=419\times 10^3\,\mbox{s}^{-1}$. At resonance, this drops to
$I_1=272\times 10^3\,\mbox{s}^{-1}$ through interference with the
field returning from within the cavity via the input mirror. This is
the base level of the graph in Fig.~\ref{fig:expt}. With the arrival
of atoms, the count rate rises to a peak of $I_2= 315\times
10^3\,\mbox{s}^{-1}$ because the field in the cavity, and hence the
field leaking back out through the input mirror, is reduced by the
presence of the atoms. At resonance, the ratio of intra-cavity
fields with no atoms and with $N$ atoms is just
$(\sqrt{I_0}-\sqrt{I_1})/(\sqrt{I_0}-\sqrt{I_2})$, which equates to
$(1+Ng^2/\kappa \gamma)$~\cite{Berman1994}. Since the Zeeman
sublevels are mixed in the optical molasses, we take
$g^2=(3/7)\,g_0^2$, where the factor of $3/7$ averages over all the
$F=3$ sublevels. The solid curve in Fig.~\ref{fig:expt} is a fit to
this model for a cloud with Gaussian density distribution, which
gives $N=0.7$ at the peak. We expect this result to be a slight
underestimate because the number of atoms is randomly fluctuating
and the reflected power is a nonlinear function of that number,
saturating at higher values. Nevertheless, it shows that $N\simeq
1$, in agreement with our knowledge of the initial atom number and
temperature in the MOT.

\begin{figure}
\includegraphics[scale=0.6]{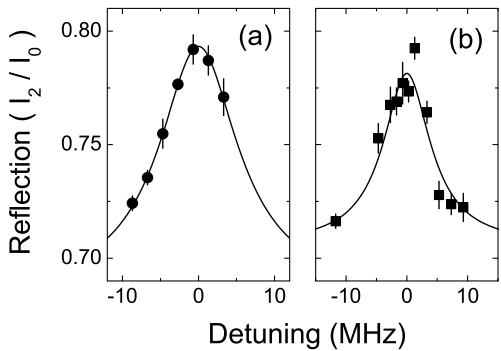}
\caption{\label{fig:spectra}Peak probe laser reflection with
$\omega_C=\omega_A$ versus detuning $(\omega_L-\omega_A)/2\pi$ for
different initial MOT numbers. (a): $\langle N \rangle =1.1$ atoms
in the cavity; full-width at half-maximum $2.1 \times 2\gamma$.
(b): $\langle N \rangle =0.6$; width $1.7 \times 2\gamma$. Solid
curves are Monte Carlo fits to the data using
Eq.~(\ref{eq:reflection}).}
\end{figure}

For a more detailed study, we have measured the fraction of power
reflected at the atom peak $I_2/I_0$ as a function of the probe
laser frequency $\omega_L$, keeping the cavity frequency $\omega_C$
centered on the atomic resonance frequency $\omega_A$. The data
points in Fig.\ \ref{fig:spectra} show the result of this scan for
two different values of the initial atom number in the MOT. The
solid lines are a fit to the functional
dependence~\cite{Thompson1992,Horak2003}
\begin{eqnarray}\label{eq:reflection}
    \frac{I_2}{I_0} = \frac{
    \left[1+\frac{Ng^2/(\kappa\gamma)}{1+(\Delta/\gamma)^2}-\beta \right]^2
    + \left(\frac{\Delta}{\kappa}\right)^2\left[1-\frac{N(g/\gamma)^2}{1+(\Delta/\gamma)^2}\right]^2}{
    \left[1+\frac{Ng^2/(\kappa\gamma)}{1+(\Delta/\gamma)^2}\right]^2
    +
    \left(\frac{\Delta}{\kappa}\right)^2\left[1-\frac{N(g/\gamma)^2}{1+(\Delta/\gamma)^2}\right]^2}\,
    ,
\end{eqnarray}
where $\Delta=\omega_L-\omega_{C,A}$ is the probe laser detuning.
The parameter $\beta$ characterizes the maximum fringe contrast,
and can be determined from the response of the resonant empty
cavity, according to $I_1/I_0=(1-\beta)^2$. For our parameters,
Eq.(\ref{eq:reflection}) describes a spectrum whose peak and width
both increase with $N$. In order to account for number
fluctuations, we fit this formula to the data using a Monte Carlo
simulation, similar to that of Ref.\ \cite{Thompson1992}, but with
two additions. First, because our Rayleigh length $z_R=w_0^2/(2k)$
is comparable to the cavity length, we include in our model the
variation of the transverse mode size $w=w_0\sqrt{1+(z/z_R)^2}$.
Second, we convolve the calculated spectra with a Lorentzian to
account for the measured linewidth of our probe laser, which is
$1.2\,\mbox{MHz}$. The Monte-Carlo fits shown in Figs.\
\ref{fig:spectra}(a) and \ref{fig:spectra}(b) yield average atom
numbers of $\langle N\rangle =1.1$ and $0.64$, with spectral
widths of $2.1\times 2\gamma$ and $1.7\times 2\gamma$,
respectively. This result demonstrates that our microcavities can
measure very small atom numbers, even at such modest finesse.

\begin{figure}
\includegraphics[scale=0.6]{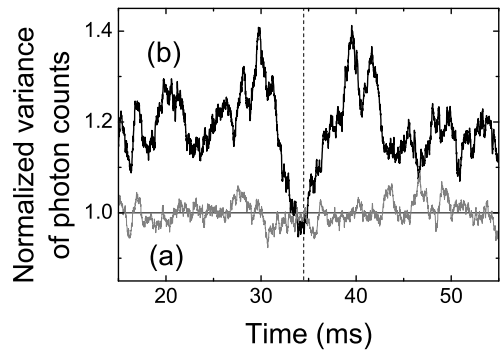}
\caption{\label{fig:subpoisson}Normalized variance of the photon
counts reflected from the cavity as a function of time, determined
from 48 drops. (a) Nonresonant cavity. (b) Resonant cavity.
Horizontal line: shot-noise limit. Vertical dashed line: arrival
time of the peak number of atoms.}
\end{figure}

Another way to detect the arrival of atoms in the cavity is through
the photon statistics. Therefore, we performed 48 drops of the atom
cloud, recording the APD signal versus time as a series of counts
$m$, each being the integral of the rate over a time interval
$\tau=10\,\mu\mbox{s}$. These counts were corrected~\cite{deadtime}
for the measured dead time of the detector $\tau_d =44\,\mbox{ns}$
according to $n=m/(1-m\,\tau_d/\tau)$. The normalized variance of
the counts $f(n)\equiv \mbox{Var}(n)/\langle n\rangle$ was then
corrected~\cite{BachorRalph} for the $90\%$ beam splitter
transmission and the $60\%$ APD quantum efficiency according to
$f_{\rm corr}(n)-1=[f(n)-1]/0.54$ in order to derive the normalized
variance of the photon numbers shown in Fig.~\ref{fig:subpoisson}.
With the cavity far from resonance and in the absence of atoms
(curve [a]), this is close to unity, as expected for statistical
noise. However, on resonance (curve [b]), there is additional
technical noise due to vibrations of the cavity length. This curve
has audio-frequency fluctuations, which are also seen as distinct
peaks in the Fourier spectrum of the noise recorded over longer time
intervals. The excess noise is suppressed by the arrival of the
atoms at a time $t\sim 35\,\mbox{ms}$. In part this is because the
reflection minimum is less sensitive to variations of cavity length
when the atoms are present. This suppression mechanism accounts for
only half the noise reduction we observe. The additional noise
reduction is of unknown origin, but may be
quantum-mechanical~\cite{Rempe1991}. This will require further
investigation after the stability of the upper cavity length has
been improved using active stabilization of the lower cavity
(Fig.~\ref{fig:expt}) during the atom drop.

\begin{figure}
\includegraphics[scale=0.7]{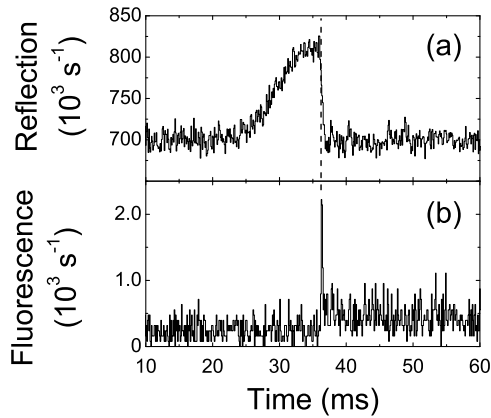}
\caption{\label{fig:fluorescence}(a) Probe reflection signal showing
abrupt atom loss when the excitation laser is turned on. (b)
Cavity-stimulated decay generates a pulse of photons in the fiber
when the atom cloud is excited. The dotted line shows coincidence of
the photon pulse with the turn-on of the excitation laser.}
\end{figure}

The radiation pattern of an atom excited in the cavity is strongly
enhanced in the direction of the cavity mode, a phenomenon known as
the Purcell effect~\cite{Purcell}. Thus the cavity enhances the
emission of photons into the single-mode fiber attached to the plane
mirror. Without a cavity, only about $10^{-4}$ of the fluorescence
photons would radiate into the fiber. By contrast, for our
parameters, the rate of emission into the cavity $2g^2/\kappa$ is
roughly equal to the free-space rate $2\gamma$.  In order to
demonstrate this effect in our microcavities, we prepare and drop an
atom cloud as described above, monitoring the reflection of the weak
probe laser to determine when the atom number is at its peak value
of $N\simeq 1$. At this point we turn on a resonant laser beam
(40\,mW/cm$^2$) propagating transverse to the cavity axis, in order
to excite the atoms.  As shown in Fig.~\ref{fig:fluorescence}(a),
the reflection signal recorded on the APD drops abruptly when the
excitation laser is turned on. This is due to a combination of
optical pumping into the dark $F=2$ state and physical removal of
the atoms from the cavity as a result of the radiation pressure. The
experiment is repeated without the probe laser, but under otherwise
identical conditions, so that the cavity-stimulated photons can be
detected using the APD. As shown in Fig.~\ref{fig:fluorescence}(b),
we see a burst of fluorescence at the moment of excitation. This is
absent if there are no atoms in the cavity or if the cavity is
detuned from resonance, confirming that the detected light is indeed
due to the Purcell-enhanced transfer of photons from the excitation
laser beam into the fiber via both the atoms and cavity. Typically
we detect one or two photons in the peak, which is consistent with
the prediction of a master equation model, taking into account
cavity-enhanced optical pumping into the dark $F=2$ state and the
number of atoms known from Fig.~\ref{fig:fluorescence}(a). The
efficiency of excitation transfer from the atom to the cavity is
limited at present by the $50\%$ of photons radiated out of the
cavity, but with the higher-quality $\mathcal{F}> 5000$ cavities
available in our laboratory, we anticipate reaching over $92\%$ in
future experiments. This would make our chip cavity a useful source
of photons on demand and the entanglement between the atom and the
photon in the fiber would be a useful resource for quantum
information processing.

In conclusion, we have used a micro-fabricated optical cavity based
on a silicon wafer to detect less than one atom on average with a
signal/noise ratio of 3 over $\sim 250\mu$s. In the next experiment
this cavity will be rebuilt to give a much improved fringe
visibility and hence a reduced detection time of $\sim 2\mu$s,
allowing us to follow a given single atom during the $\sim 15\mu$s
it takes to fall through the cavity. Our cavity design results in a
large single-photon Rabi frequency, and cavity photons are
conveniently coupled to a single-mode optical fiber or waveguide,
which is integral to the construction.  Cold atoms falling through a
cavity were detected by their effect on the cavity reflection
spectrum and by their influence on the noise in the reflected light.
We also triggered the emission of photons into the fiber by means of
the Purcell effect. These results constitute first steps towards
building an optical micro-cavity network on a chip for applications
in quantum information processing.

\begin{acknowledgments}
This work was supported by EU networks Atom Chips, Conquest, and
SCALA,  by the Royal Society, and by EPSRC grants for QIPIRC, CCM
programme and Basic Technology. The group of M. Kraft at Southampton
University etched the silicon mirrors.
\end{acknowledgments}


\begin{thebibliography}{10}

\bibitem{Berman1994}
\textit{Cavity Quantum Electrodynamics}, edited by P. R. Berman
(Academic Press, San Diego, 1994).

\bibitem{vanEnk2004}
S. J. van Enk, H. J. Kimble, and H. Mabuchi, Quantum Inf. Process.
{\bf 3}, 75 (2004).

\bibitem{Kuhn2002}
A. Kuhn, M. Hennrich, and G. Rempe, Phys. Rev. Lett. {\bf 89},
067901 (2002)

\bibitem{McKeever2004}
J. McKeever {\it et al.}, Science {\bf 303}, 1992 (2004); M.
Hijlkema {\it et al.}, Nature Physics {\bf 3}, 253 (2007).

\bibitem{Vuletic}
A. T. Black, J. K. Thompson, and V. Vuleti\'{c}, Phys. Rev. Lett.
{\bf 95}, 133601 (2005); A. T. Black, J. K. Thompson, and V.
Vuleti\'{c}, Science {\bf 313}, 74 (2006).

\bibitem{Cirac1997}
J. I. Cirac, P. Zoller, H. J. Kimble, and H. Mabuchi, Phys. Rev.
Lett. {\bf 78}, 3221 (1997).

\bibitem{DiVincenzo2000}
D. P. DiVincenzo, Fortschr. Phys. {\bf 48}, 771 (2000).

\bibitem{clusters}
R. Raussendorf and H. J. Briegel, Phys. Rev. Lett. {\bf 86}, 5188
(2001).

\bibitem{multicavity}
L. M. Duan and H. J. Kimble, Phys. Rev. Lett. {\bf 90}, 253601
(2003); D. E. Browne, M. B. Plenio, and S. F. Huelga, {\it ibid.}
{\bf 91}, 067901 (2003); S. Clark, A. Peng, M. Gu, and S. Parkins,
{\it ibid.} {\bf 91}, 177901 (2003); M. J. Hartmann, F. G. S. L.
Brandao, and M. B. Plenio, Nature Phys. {\bf 2}, 849 (2006); A. D.
Greentree, C. Tahan, J. H. Cole, and L. L. C. Hollenberg, {\it
ibid.} {\bf 2}, 856 (2006); A. Serafini, S. Mancini, and S. Bose,
Phys. Rev. Lett. {\bf 96}, 010503 (2006); Z. Yin and F. Li, Phys.
Rev. A {\bf 75}, 012324 (2007).

\bibitem{Teper2006}
I. Teper, Y.-J. Lin, and V. Vuleti\'{c}, Phys. Rev. Lett. {\bf 97}
023002 (2006); A. Haase, B. Hessmo and J. Schmiedmayer, Opt. Lett.
{\bf 31}, 268 (2006).


\bibitem{Aoki2006}
T. Aoki {\it et al.}, Nature (London) {\bf 443}, 671 (2006).

\bibitem{Steinmetz2006}
T. Steinmetz {\it et al.}, Appl. Phys. Lett. {\bf 89}, 111110
(2006).

\bibitem{Trupke2005}
M. Trupke {\it et al.} Appl. Phys. Lett. {\bf 87}, 211106 (2005).

\bibitem{Eriksson2005}
S. Eriksson {\it et al.}, Eur. Phys. J. D {\bf 35}, 135 (2005).

\bibitem{microoptics}
See, for example, A. A. Bettiol {\it et al.}, Nucl. Instr. and Meth.
in Phys. Res. B {\bf 31}, 364 (2005).

\bibitem{Gollasch2005}
C. O. Gollasch {\it et al.}, J. Micromech. Microeng. {\bf 15}, S39
(2005).

\bibitem{Volz1996}
U. Volz and H. Schmoranzer, Phys. Scr. {\bf T65}, 48 (1996).

\bibitem{SalehTeich}
B. E. A. Saleh and M. C. Teich, {\it Fundamentals of Photonics}
(John Wiley \& Sons, New York, 1991).

\bibitem{Purcell}
E. M. Purcell, Phys. Rev. {\bf 69}, 681 (1946); D. Kleppner, Phys.
Rev. Lett. {\bf 47}, 233 (1981).

\bibitem{subdoppler}
P. D. Lett {\it et al.}, Phys. Rev. Lett. {\bf 61}, 169 (1988).

\bibitem{Horak2003}
P. Horak {\it et al.}, Phys. Rev. A {\bf 67}, 043806 (2003).

\bibitem{Thompson1992}
R. J. Thompson, G. Rempe, and H. J. Kimble, Phys. Rev. Lett. {\bf
68}, 1132 (1992).

\bibitem{deadtime}
See, for example, D. F. Yu and J. A. Fessler, Phys. Med. Biol. {\bf
45}, 2043 (2000).

\bibitem{BachorRalph}
H.-A. Bachor and T. C. Ralph, {\it A Guide to Experiments in
Quantum Optics} (Wiley VCH, Weinheim, 2004).

\bibitem{Rempe1991}
G. Rempe {\it et al.}, Phys. Rev. Lett. {\bf 67}, 1727 (1991); see
also H. J. Carmichael {\it et al.}, Opt. Commun. {\bf 82}, 73
(1991).

\end{thebibliography}
\end{document}